\documentstyle{article}
\begin{document}
\def\aproxgt{\mathrel{%
      \rlap{\raise 0.511ex \hbox{$>$}}{\lower 0.511ex \hbox{$\sim$}}}}
\def\aproxlt{\mathrel{%
      \rlap{\raise 0.511ex \hbox{$<$}}{\lower 0.511ex \hbox{$\sim$}}}}
\title{Relativistic Diskoseismology}
\author{Robert V. Wagoner\\
{\it Dept.~of Physics and Center for Space Science \& Astrophysics}\\
{\it Stanford University, Stanford, CA 94305--4060 USA}\\
{\it wagoner@leland.stanford.edu}\\}
\date{}
\maketitle

\begin{abstract}

We will summarize results of calculations of the modes of oscillation trapped
within the inner region of accretion disks by the strong-field gravitational
properties of a black hole (or a compact, weakly-magnetized neutron
star). Their driving and damping will also be addressed. The focus will be on
the most observable class: the analogue of internal gravity modes in
stars. Their frequencies which corrrespond to the lowest mode numbers depend
almost entirely upon only the mass and angular momentum of the black
hole. Such a feature may have been detected in the X-ray power spectra of two
galactic `microquasars', allowing the angular momentum of the black hole to be
determined in one case.

\end{abstract}

\noindent
\leftline{PACS: 04.70.Bw; 97.60.Lf}
\leftline{Keywords: Black holes; Accretion disks} 

\section{Introduction}

   In this review, dedicated to one of my oldest and best friends, Giora Shaviv, on the occasion of his 60th birthday, a potentially powerful probe of black holes will be described. But first we may ask to what extent their existence has been verified, keeping in mind our definition of a black hole: a region of spacetime described by the Kerr metric. There are several lines of evidence. 

A) The determinations that a certain amount of mass is contained within a certain radius (typically many orders of magnitude greater than the horizon size of a black hole of that mass) are based mainly upon (a) the time dependence of the Doppler velocity of a companion for binary systems and (b) the Doppler velocity and orbital radius of gas or stars for galactic nuclei. There is strong evidence for a mass greater than that of standard neutron stars in roughly six Galactic binaries (McClintock 1998), and of supermassive objects in roughly twice as many galactic nuclei. The rapid variability and high energies of much of the radiation provides evidence that the emission region is indeed compact ($GM/Rc^2\aproxgt 0.1$).

B) The energy dependence of emission line profiles, when interpreted as Doppler and gravitational shifts of excited line emission from accretion disks, provides values of its extent, in terms of $R/M$ (e.~g., Fabian et al.~1989; Tanaka et al.~1995). However, a dependence of the line emissivity on radius must be assumed. In one case, it has been claimed that a rapidly rotating black hole ($a\equiv cJ/GM^2\approx 1$ is required to explain inferred values of inner radius $r_i < 2GM/c^2$ (Iwasawa et al.~1996).

C) Another approach involves fits to the energy dependence of the continuum (X-ray) spectrum (Zhang et al.~1997). A disk luminosity and maximum temperature is obtained for sources of known distance in which the non power-law component of the energy spectrum can be fit with a standard (diluted blackbody) optically thick accretion disk model. This provides an effective area. After radiative transfer and general relativistic corrections, an inner radius of the disk is extracted. For luminosities sufficiently below the maximum (Eddington luminosity), $r_i$ is mainly a function of $M$ and $a$.

D) Recently, Narayan et al.~(1997,1998) have claimed evidence for an event horizon, based upon the advectively dominated class of accretion disk models that have been developed (by Abramowicz, Lasota, Narayan, and others). These flows carry most of the energy generated into the black hole, rather than being radiated. Without a horizon, the energy would appear when the flow hit the surface of the central object (such as occurs with a neutron star). From spectral fits and the determination of the range of luminosities of systems whose accretion rate varies, they have identified various Galactic X-ray black hole candidates, as well as our and other galactic nuclei, as being in this state. An accretion rate less than about $8 \%$ of the maximum is required. However, there is still no criterion to determine whether a disk with such an accretion rate will instead choose to be in the standard (radiating) state. The low luminosity would then be due to a lower accretion rate.  

E) The frequencies of narrow features in the power spectra of various black hole candidates have been ascribed to persistent structures in the associated accretion disk. There have been various proposals for an identification with the orbital frequency of hot `blobs', usually at the innermost radius of the disk. However, it is not clear (a) how such `blobs' form, (b) how they survive as coherent structures, and (c) why they are confined to a particular radius. Another identification is with inertial-acoustic traveling waves, which we shall comment upon later. Instead, we focus on the spectrum of normal modes of oscillation, which must exist at some level (determined by the driving and damping processes in the disk). In the same spirit with which helioseismology probes the interior of the Sun, this probe of the Kerr metric (and its accretion disk) has been dubbed (relativistic) diskoseismology. We now analyze this approach. [For an up-to-date survey of black hole accretion disk theory, see the monograph by Kato, Fukue, and Mineshiga (1998).]

Since 1990, our group has been investigating consequences of the realization [initially by Kato and Fukue (1980)] that general relativity can trap normal modes of oscillation near the inner edge of accretion disks around black holes. The strong gravitational fields that are required can also be produced by neutron stars that are sufficiently compact (requiring a soft equation of state) and weakly magnetized that there is a gap between the surface of the star and the innermost stable orbit of the accretion disk. Although we shall not explicitly consider such neutron stars here, the results obtained will also apply to them to first order in the dimensionless angular momentum parameter $a=cJ/GM^2$, since their exterior metric is
identical to that of a black hole to that order. It should be noted that $a\aproxlt 0.5$ for almost all models of rotating neutron stars (Friedman \& Ipser 1992; Cook, Shapiro, \& Teukolsky 1994).

These modes of oscillation provide a potentially powerful probe of both
strong gravitational fields and the physics of accretion disks, since:

$\bullet$ They do not exist in Newtonian gravity

$\bullet$ Their frequencies depend upon the angular momentum as well as the mass of the black hole

$\bullet$ The fractional frequency spread of each mode depends upon $\alpha$, the elusive viscosity parameter of the accretion disk.

\section{Adiabatic Normal Mode Oscillations}

   Following exploratory calculations by Kato \& Fukue (1980), Okasaki et al.
(1987), and Kato (1989), we also employed a modified Newtonian potential to
calculate the adiabatic eigenfunctions and eigenvalues of the lowest
acoustic ($p$) modes (Nowak \& Wagoner 1991) and internal gravity ($g$)
modes (Nowak \& Wagoner 1992, 1993). This was extended to full general
relativity by Perez (1993), who also included the corrugation ($c$) modes
studied by Kato (1990, 1993) and Ipser (1994, 1996). The g--modes have been
analyzed more extensively by Perez, Silbergleit, Wagoner, \& Lehr (1997),
and we shall also summarize their key results below.

We have applied the general relativistic perfect fluid perturbation
formalism of Ipser and Lindblom (1992) to thin accretion disks in the Kerr
metric. The radial and vertical components of the velocity of the fluid is neglected,
which has little effect on the modes for these thin models. Neglecting the
gravitational field of the disk (also a good approximation), the adiabatic
oscillations of all physical quantities can be expressed in terms of a
single scalar potential $\delta V(r,z)\propto \delta p/\rho$, governed by a second-order partial
differential equation. The stationary ($\partial /\partial t=0$) and
axisymmetric ($\partial /\partial\phi=0$) unperturbed accretion disk is
taken to be described by the standard relativistic $\alpha$-disk model (Novikov and
Thorne 1973, Page and Thorne 1974). We employ gravitational units ($G=c=1$).

All fluid perturbations are of the form $f(r,z)\exp[i(\sigma t + m\phi)]$.
With $\Omega$ the fluid angular velocity, the comoving frequency is
$\omega(r) = \sigma + m\Omega(r)$. It is sufficient to consider
eigenfrequencies $\sigma < 0$ and axial mode integers $m \ge 0$.
Nonaxisymmetric modes ($m\ne 0$) should produce relatively little luminosity
modulation, unless the disk is viewed close to edge-on. In general, the
vertical extent of the modes within the disk is restricted by the
requirement that $|\omega|$ be greater than the buoyancy frequency.
Numerical results have been obtained for accretion disks which are
barotropic (e.~g., isentropic) on scales of order their thickness, in which
case this restriction is not operative since the buoyancy frequency
vanishes. 

The effective radial wavelengths $\lambda_r$ are significantly smaller than $r$,
allowing a WKB expansion of the problem and approximate separation of the governing equations, with $\delta V = V_r(r)V_\eta(r,\eta)$. The vertical separability coordinate is $\eta = z/h(r)$, with $h$ the effective half thickness of the disk. It and $V_\eta$ are slowly varying (compared to $V_r$) functions of $r$. 

The radial dependence of the fluid perturbations are governed by the equation
\begin{equation}
 \frac{d^2W}{dr^2} + \frac{g_{rr}(U^t)^2}{c_s^2(r,0)}
\left[\left(\frac{\Omega_\perp}{\omega}\right)^2\Psi - 1\right] (\kappa^2-
\omega^2)W = 0     \; ,  \label{W}
\end{equation}
where $g_{rr}$ is a Kerr metric component, $U^t$ is the fluid four-velocity
component $dt/d\tau$, $c_s(r,z)$ is the speed of sound, and $\Psi(r)$ is
the slowly-varying separation function. The eigenfunction $W(r)= (\kappa^2-\omega^2)^{-1}dV_r/dr$ is proportional to the radial component of the fluid displacement. 
The key controlling ingredient in this problem is the relativistic behavior of the radial epicyclic frequency $\kappa(r)$, shown in Figure~\ref{orbital} with other important orbital frequencies. As usual, $\kappa$ is the frequency of radial perturbations of circular orbits of a
free particle, while $\Omega_\perp$ is the same for vertical perturbations.

From equation (\ref{W}), we can see where the modes are trapped ($W^{-1}d^2W/dr^2<0$). Their classification is governed by the value of the eigenvalue $\Psi$, as follows.

\medskip
$\bullet$ The g (inertial-gravity)--modes [defined by $\Psi > (\omega/\Omega_\perp)^2$] are trapped where $\omega^2 < \kappa^2$. This occurs between some radii $r_-(\sigma)$
and $r_+(\sigma)$. The lowest modes (fewest number of radial and
vertical nodes in their eigenfunctions) have approximately vertical
displacements, and have eigenfrequencies $|\sigma|=2\pi f$ which are close to the
maximum possible, $\sigma_m = \kappa(r_m) + m\Omega(r_m)$. The dependence of this frequency on black hole angular momentum is plotted in Figure~\ref{freqs}, for $m=0$. Plotted in Figure~\ref{size} is this radius $r_m$ to which $r_-$ and $r_+$ converge as the frequency is raised to this
maximum, as well as the effective radial width ($r_+ - r_-$) of the eigenfunction. Although the approximate center of the eigenfunction (located at $r_m$) is close to the temperature maximum of the disk, its relatively small width appears to limit the luminosity modulation to at
most a few percent.

The lowest radial ($m=0$) g--modes have a frequency given by
\begin{equation}
 f=714(1-\epsilon_{nj})(M_\odot/M)F(a) \;\; \mbox{Hz} \; , \quad\quad
\epsilon_{nj} \approx \left(\frac{n+1/2}{j+\delta}\right)\frac{h}{r_m} \; .
\label{f} \end{equation}
The properties of the accretion disk enter only through the small mode-dependent term $\epsilon_{nj}$, which involves the disk thickness 2$h(r_m)$
and the radial ($n$) and vertical ($j$) mode numbers, with $\delta\sim1$.
Typically $h(r_m)/r_m \sim 0.1L/L_{Edd}$ for a radiation-pressure dominated
optically thick disk region. $F(a)$ is the upper function plotted in
Figure~\ref{maxfreq}. From Figure~\ref{maxfreq}, we also see that the
higher axial modes have a somewhat different dependence upon the angular
momentum of the black hole, which in principle would allow its
determination as well as that of the mass.

\medskip
$\bullet$ The p (inertial-acoustic)--modes [defined by $\Psi < (\omega/\Omega_\perp)^2$] are trapped
where $\omega^2 > \kappa^2$. The radial ($m=0$) p--modes that are trapped
between the inner radius of the disk ($r_i$) and $r_-$ have very little
radial extent, and thus will produce relatively little direct luminosity
modulation, although they will modulate the accretion onto the black hole
(or neutron star). The highest frequency modes that are trapped within
$r_+$ and the outer radius of the disk ($r_o$) will modulate a significant
fraction of the disk where the luminosity per unit radius is highest.
However, since their wavelength $\lambda(r) \sim 2\pi c_s/\omega$ will be
relatively short, their damping will be enhanced and the net luminosity modulation will be reduced. We have not yet investigated this outer branch of p--modes because of the
uncertain physics and location of the outer radius.

It should be noted that Honma, Matsumoto, \& Kato (1992) and Milsom \& Taam (1997) have claimed that inertial acoustic waves can be generated in disks with sufficient viscosity and fairly low mass accretion rates. These axisymmetric waves appear to propagate from the radius $r_m$ where $\kappa$ achieves its maximum, and have the corresponding frequency of the fundamental g--mode. However, the dependence of their wavelength $\lambda_r=2\pi/k_r$ on radius does not agree with the simple dispersion relation they assumed ($\omega^2=\kappa^2 + k_r^2c_s^2$). Possibly g--mode oscillations have coupled into these outgoing waves.  

\medskip
$\bullet$ The c (corrugation)--modes [defined by $\Psi \cong (\omega/\Omega_\perp)^2$] are
typically nonradial ($m=1$) vertically incompressible waves near the inner edge of the disk that slowly precess around the angular momentum of the black hole. The changing projected area of the mode could modulate the luminosity. The eigenfrequency of the fundamental mode and the corresponding trapping radius $r_c$ are given by 
\begin{equation}
2\pi f = |\sigma|= \Omega(r_c) - \Omega_\perp(r_c) \; , \quad 
r_c \cong r_i + CM(a^{-1/3}-1) \; . \label{prec}
\end{equation}
The mode extends from $r_i$ to $r_c$. The constant $C$ depends upon the speed of sound in the disk; a typical value is $C=0.17$. Thus the mode has significant radial extent only for $a\ll 1$, in which case the frequency becomes small and equal to the Lense-Thirring value: $|\sigma|\rightarrow 2aM^2/r_c^3$. The dependence of the frequency $f$ on black hole angular momentum is also plotted in Figure~\ref{freqs}.

The frequencies of all of these modes are proportional to $1/M$, but their dependence on the angular momentum of the black hole is quite different (c.~f.~Perez 1993). It appears that the 
g--modes are the most robust, however, because their trapping does not involve the uncertain dissipative properties of the inner or outer disk radius. They should also be the most observable, because they usually occupy the largest area of the disk, near the temperature maximum.

Also plotted in Figure~\ref{freqs} is the widely invoked frequency $\Omega_{max}/2\pi$ of a `blob' which orbits at the inner radius $r_i$ of the disk. Note that the ratio of this frequency to that of the fundamental g--mode remains fairly close to 3.08 (its value for $a=0$) at larger values of $a$.

\section{Oscillation Amplitude and Frequency Width}

Nowak and Wagoner (1992) have estimated the damping (or growth) rates of p
and g--modes due to isotropic and anisotropic turbulent viscosity and
gravitational radiation reaction. Although the gravitational radiation
instability will cause the modes to grow for $M \gg 10^8M_\odot$, we
have found that isotropic viscosity will always produce damping.
With $j$ and $n$ of order the number of vertical and radial nodes in any
particular eigenfunction, its corresponding quality factor $Q$ and 
frequency broadening is given by
\begin{equation}
1/Q_{jn} \approx (\Delta f/f)_{jn} \sim [j^2 + (h/r)n^2]\alpha \; ,
\label{Q}
\end{equation}
where the viscosity parameter is thought to lie in the range $10^{-
2} < \alpha < 1$. However, the effective width of the g--mode feature 
will also be determined by what range of mode numbers $j$ and $n$ 
are sufficiently excited.

   An estimation of the amplitude of the oscillations, and the resulting
fractional modulation of the disk luminosity, was also obtained by Nowak
and Wagoner (1993). They found approximate equipartition between the energy
of the lowest modes and the energy of the largest turbulent eddies. The turbulence can
be driven by the Velikov--Chandrasekhar--Balbus--Hawley magnetic
instability [for relevant numerical results, see Stone et al.~(1996)]. This
produces maximum $g$-mode displacements of order the thickness of the disk,
which should produce maximum luminosity modulations $\delta L_\nu /L_\nu
\sim 10^{-2}$, if the photon frequency $\nu > h^{-1}kT(max)$.

   If turbulence drives these oscillations, it should also contribute to
the fluctuation power in a broad range of frequencies. Nowak and Wagoner
(1995) have computed the resulting power spectral density, and found that
it could match that of the best-observed black hole candidate Nova Muscae
(Miyamoto et al.~1993), but only at frequencies $f > 10$ Hz. The
power is predicted to drop rapidly ($\propto f^{-5}$) above the frequency
$\Omega_{max}/2\pi$.

Another potential excitation mechanism is `negative radiation damping', produced by radiative losses in a radiation-pressure dominated region of the disk (Nowak et al.~1997). This can be understood from the dispersion relation for our modes, obtained from equation (\ref{W}) in the extreme WKB limit:
\begin{equation}
\omega^2 \approx \kappa^2 \pm c_s^2k_r^2 \; , \quad (\mbox{p--mode: +, g--mode: -}) \; .
\label{dispersion}
\end{equation}
This expression reflects the fact that the pressure forces work against the effective radial gravity for the g--modes, so the loss of pressure stengthens the mode. Again, the growth rate is proportional to $\alpha$.

\section{Observations}

   These modes of oscillation should modulate the quasi-thermal emission
from accretion disks in the ultraviolet for supermassive black holes and at
soft x-ray energies for stellar mass black holes. We now discuss two black-hole candidate binary X-ray sources whose power spectra (obtained with the RXTE satellite) contained a feature which could be produced by a g--mode. The strongest evidence is the fact that (unlike lower frequency features that have been seen in various black hole candidates) the frequency did not change as the source luminosity (and presumably mass accretion rate) did. This lack of dependence on the properties of the accretion disk is clearly predicted by equation (\ref{f}) for the lowest g--mode states. It is intriguing that these two sources are the strongest `microquasars' in our galaxy.

Morgan, Remillard, and Greiner (1996) have detected a feature at $f=67$ Hz with
$\Delta f/f \approx 0.05$ in the microquasar GRS 1915+105. During the two periods that the feature was detectable, the source luminosity varied by a factor of $\sim 2$ while the frequency changed by less than 3\%. The amplitude of the peak varied from 1.5\% at 3 keV to 6\% at 15 keV. If this feature is produced by a g--mode oscillation, equation (\ref{f})
predicts a black hole mass of $10.6 M_\odot$ if it is nonrotating to $36.3
M_\odot$ if it is maximally rotating. Other aspects of this identification
are explored by Nowak et al.~(1997).

Remillard et al.~(1997) have detected a feature at $f=300$ Hz with $\Delta f/f \approx 0.4$ in the microquasar GRO J1655-40. It was only detected during 7 observations that showed the hardest X-ray spectra. Although the feature was clearly identified (with an amplitide of 0.8\%) only by summing all observations, its frequency appeared to be stable. From extensive optical observations of the companion star, Orosz and Bailyn (1997) have determined the black hole candidate mass to be $7.02\pm 0.22 M_\odot$, and the inclination of the binary to be $70^\circ$. This agrees with the estimated direction of the jet (if it is perpendicular to the orbital plane and the accretion disk) in both microquasars. 

In Figure~\ref{J1655-40}, we plot the relations between black hole angular momentum and mass if we identify the feature as being produced by the fundamental g--mode, c--mode, or `blob' at the inner disk radius. Using a distance of 3.2 kpc and the observed energy spectrum of this source, Zhang, Cui, \& Chen (1997) obtained an inner disk radius of approximately 20 km (by the method mentioned earlier). This yields the other relation between $a$ and $M$ shown. The value of the mass determined by Orosz \& Bailyn (1997) is also indicated. 

Note that only by identifying the frequency with that of the g--mode is agreement with the spectroscopic determination attained, leading to a black hole angular momentum approximately 93\% of its maximum possible value. If further tests confirm this identification, it will be the first accurate determination of the angular momentum of a black hole. 

One apparent problem is the fact that in both microquasars, the modulated flux has a far harder spectrum than that of the 1--2 keV blackbody emission from the disk. However, the g--mode modulated luminosity will come from a radius significantly less than the average radius where the total luminosity is emitted. Consider then the effect of an atmosphere (or `corona') of very hot electrons [temperature $T_a(r)$] which scatters the photons emitted from radius $r$ $N(r)$ times. If the Compton parameter $y(\propto T_aN)$ is less than unity and increases as $r$ decreases (as seems reasonable), then the Compton scattering can significantly harden the spectrum of the modulated photons relative to the total. Reasonable limits on the optical depth will prevent the photons from becoming demodulated by the electron scattering.

\section{Conclusions}

Returning to our initial question, we have seen that the evidence for compact, nonstellar masses is strong, while the evidence for horizons is less strong but intriguing. The ultimate evidence of black holes (the Kerr metric) and measurement of their only other property (angular momentum) is still elusive. However, we have seen that the `diskoseismic' probe, as well as spectroscopic (line and continuum) methods, have great promise. For instance, we note the fact that the coincidence between the g--mode and Stefan-Boltzmann determinations of the GRO J1655-40 black hole angular momentum would have disappeared if the metric was significantly different from Kerr.

It is clear that continued X-ray timing observations at the times scales sufficiently short to  see the high frequency modes are crucial to their identification. These are being carried out by RXTE (although power spectra to higher frequency are critical) and will soon be carried out by the USA satellite. In addition, we should not lose sight of the fact that supermassive black hole accretion disks should exhibit similar signatures, merely scaled by the mass. For instance, the g--mode period for a $10^8$ solar mass slowly rotating black hole is 1.6 days. Long-term monitoring with at least 1\% accuracy photometry would be required. 

\medskip
Most of the results reported here were obtained in collaboration
with Dana Lehr, Michael Nowak, and Alexander Silbergleit.
The research was supported in part by NASA through ATP grant NAG 5-3102 to
R.V.W. and grant NAS8-39225 to Gravity Probe B.

\section*{References}

\noindent
Cook, G.B., Shapiro, S.L., \& Teukolsky, S.A. 1994, Ap.~J. 424, 823.\\
Fabian, A.C., Rees, M.J., Stella, L., \& White, N.E. 1989, MNRAS 238, 729.\\
Friedman, J.L. \& Ipser, J.R. 1992, Phil. Trans. R. Soc. London A 340, 391.\\
Honma, F., Matsumoto, R., \& Kato, S. 1992, PASJ 44, 529.\\
Ipser, J.R. 1994, Ap.~J. 435, 767.\\
Ipser, J.R. 1996, Ap.~J. 458, 508.\\
Ipser, J.R. \& Lindblom, L. 1992, Ap.~J. 389, 392.\\
Iwasawa, K., et al.~1996, MNRAS 282, 1038.\\
Kato, S. 1989, PASJ 41, 745.\\
Kato, S. 1990, PASJ 42, 99.\\
Kato, S. 1993, PASJ 45, 219.\\
Kato, S. \& Fukue, J. 1980, PASJ 32, 377.\\
Kato, S., Fukue, J., \& Mineshiga, S. 1998, {\it Black-hole Accretion Disks}, Kyoto 

University Press, Japan.\\ 
McClintock, J.E. 1998, in {\it Accretion Processes in Astrophysical Systems},

Proc. of the 8th Annual Maryland Astrophysics Conference, in press.\\
Milsom, J.A. \& Taam, R.E. 1997, MNRAS 286, 358.\\
Miyamoto, S., et al.~1993, Ap.~J. 403, L39.\\
Morgan, E.H., Remillard, R.A., \& Greiner, J. 1996, Ap.~J. 473, L107.\\
Narayan, R., Garcia, M.R., \& McClintock, J.E. 1997, Ap.~J. 478, L79.\\
Narayan, R., Mahadevan, R., \& Quataert, E. 1998, in {\it The Theory of Black Hole 

Accretion Disks}, ed. M.A. Abramowicz, G. Bjornsson, \& J. Pringle, in press.\\
Nowak, M.A. \& Wagoner, R.V. 1991, Ap.~J. 378, 656.\\
Nowak, M.A. \& Wagoner, R.V. 1992, Ap.~J. 393, 697.\\
Nowak, M.A. \& Wagoner, R.V. 1993, Ap.~J. 418, 187.\\
Nowak, M.A. \& Wagoner, R.V. 1995, MNRAS 274, 37.\\
Nowak, M.A., Wagoner, R.V., Begelman, M.C., \& Lehr, D.E. 1997, 

Ap.~J. 477, L91.\\ 
Novikov, I.D. \& Thorne, K.S. 1973, in {\it Black Holes}, ed.~C. DeWitt \& B.S. 

DeWitt, Gordon \& Breach, New York.\\
Okazaki, A.T., Kato, S., \& Fukue, J. 1987, PASJ 39, 457.\\
Orosz, J.A. \& Bailyn, C.D. 1997, Ap.~J. 477, 876.\\
Page, D.N. \& Thorne, K.S. 1974, Ap.~J. 191, 499.\\
Perez, C.A. 1993, Ph.~D. thesis, Stanford University.\\
Perez, C.A., Silbergleit, A.S., Wagoner, R.V., \& Lehr, D.E. 1997, 

Ap.~J. 476, 589.\\
Remillard, R.A. et al.~1997, in {\it Proceedings of the 18th Texas Symposium on 

Relativistic Astrophysics}, World Scientific, in press.\\
Stone, J.M., Hawley, J.F., Gammie, C.F., \& Balbus, S.A. 1996, Ap.~J. 463, 656.\\
Tanaka, Y., et al.~1995, Nature 375, 659.\\
Zhang, S.N., Cui, W., \& Chen, W. 1997, Ap.~J. 482, L155.\\

\newpage

\begin{figure}
\caption{The radial dependence of the square of the fundamental 
free-particle frequencies that govern the modes of the disk: Keplerian
($\Omega$), and radial ($\kappa$) and vertical ($\Omega_\perp$) epicyclic.
Three values of the black hole angular momentum parameter $a=cJ/GM^2$ are
chosen.}
\label{orbital}
\end{figure}

\begin{figure}
\caption{The dependence of three characteristic disk frequencies on the dimensionless angular momentum of the black hole.}
\label{freqs}
\end{figure}

\begin{figure}
\caption{(a) The black hole angular momentum dependence of the radius $r_m$
to which $r_-$ and $r_+$ converge as the g--mode eigenfrequency $|\sigma| \rightarrow \sigma_m$. Also shown is the radius $r_i$ of the inner edge of the disk.
(b) The dependence of the fractional effective width of the lowest
eigenfunction, $\Delta r/r_m = [r_+(\sigma) - r_-(\sigma)]/r_m$, on the
angular momentum of the black hole. The same values of $m$ are chosen as in
(a). The accretion disk model has a barotropic equation of state, and a speed of sound corresponding to a luminosity $L=0.1L_{Edd}$ from a radiation-pressure dominated disk.}
\label{size}
\end{figure}

\begin{figure}
\caption{(a) The dependence of the maximum radial ($m=0$) g--mode eigenfrequency on
the black hole angular momentum parameter $a=cJ/GM^2$, relative to its
value at $a=0$.
(b) The ratio of the maximum eigenfrequency of some higher $m$ modes to
that of the radial mode.}
\label{maxfreq}
\end{figure}

\begin{figure}
\caption{Relations between black hole angular momentum and mass for the microquasar GRO J1655--40. The three frequencies indicated are identified with the 300 Hz feature in its power spectrum. The value of inner radius was obtained by the spectroscopic (Stefan-Boltzmann) method of Zhang, Cui, \& Chen (1997). The black hole mass, as determined by Orosz and Bailyn (1997), is also indicated.}
\label{J1655-40}
\end{figure}

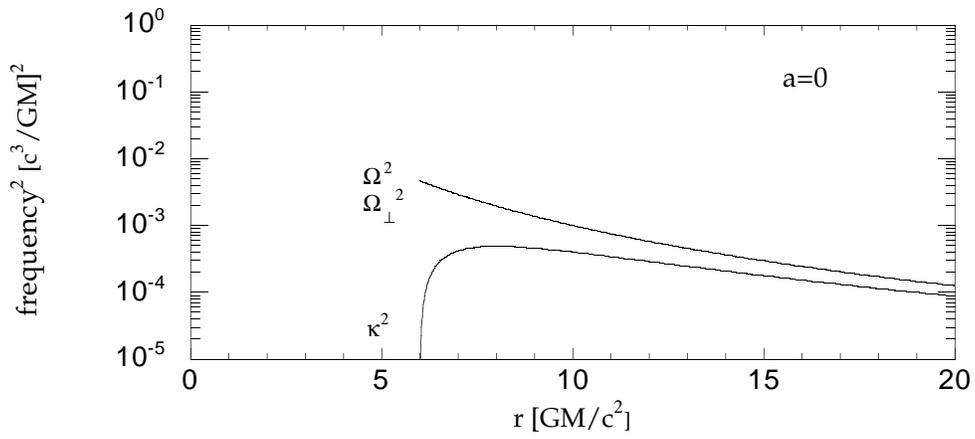
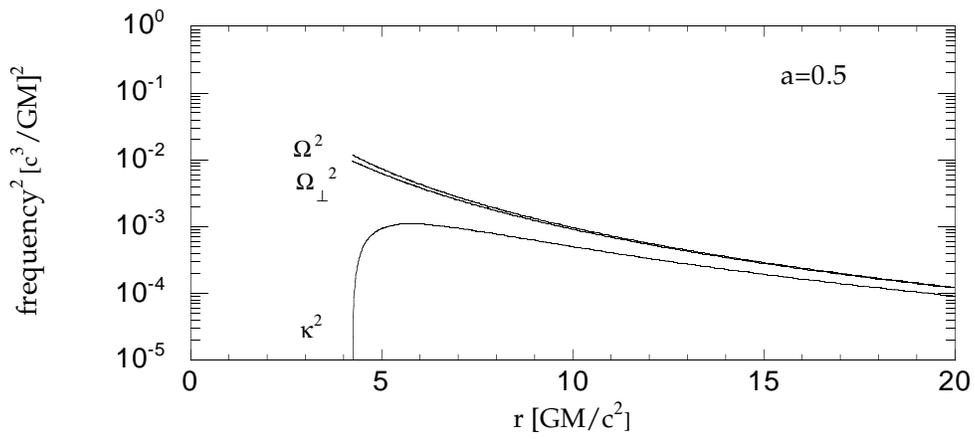
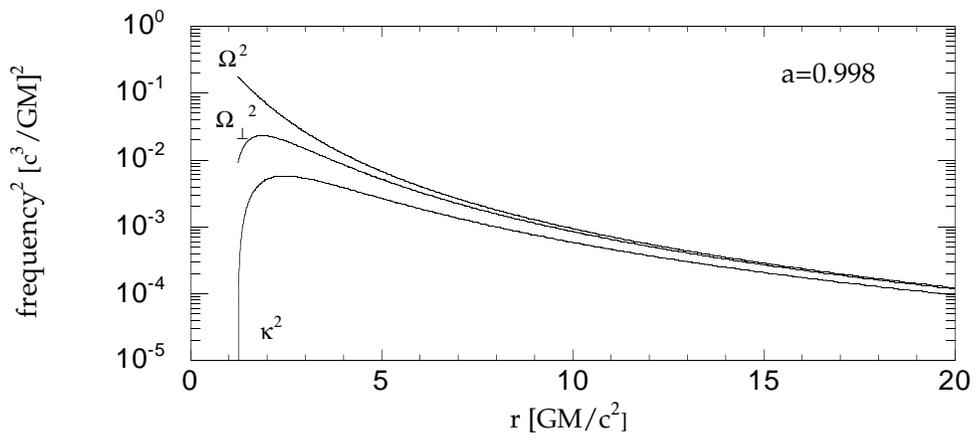
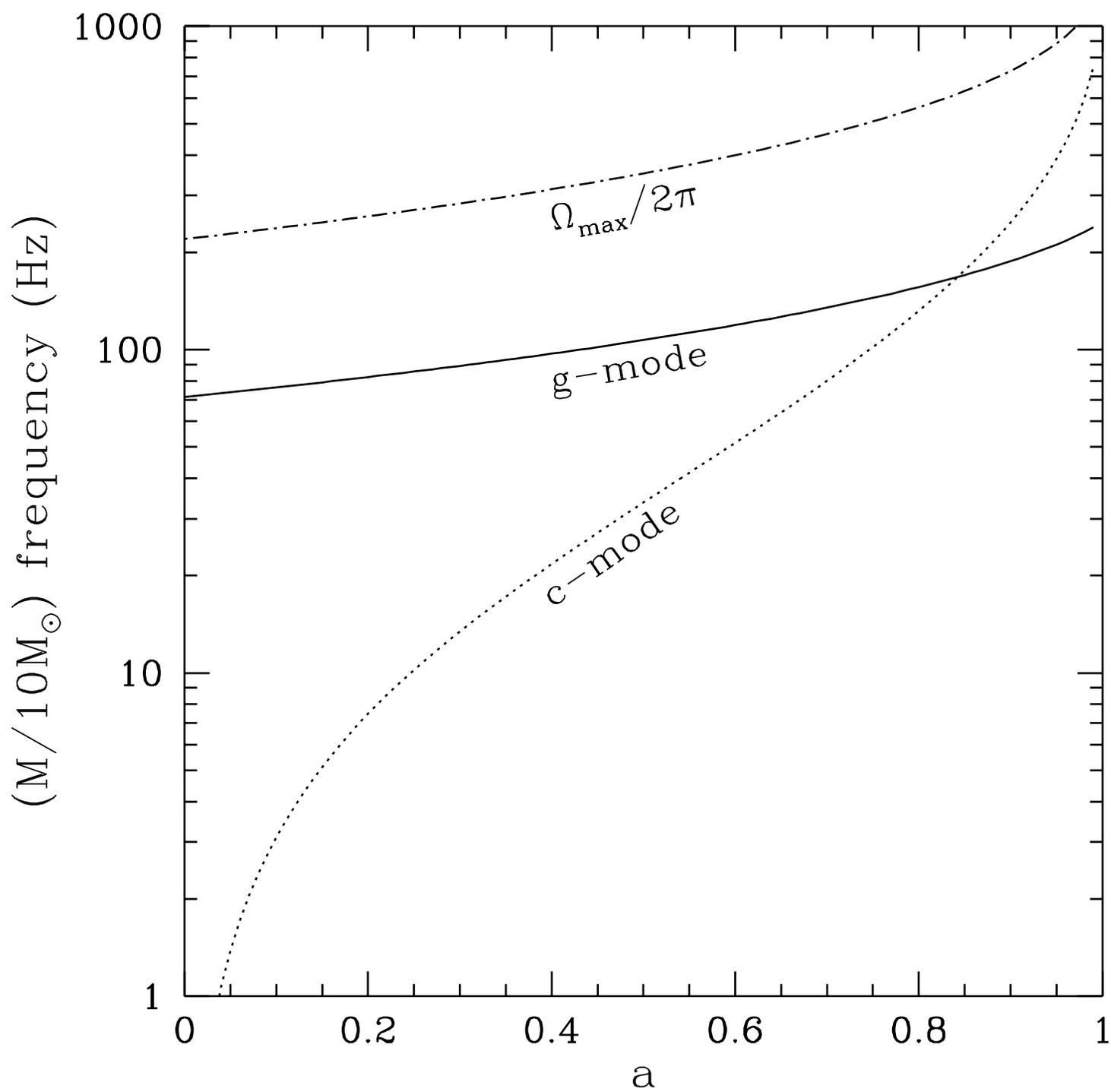
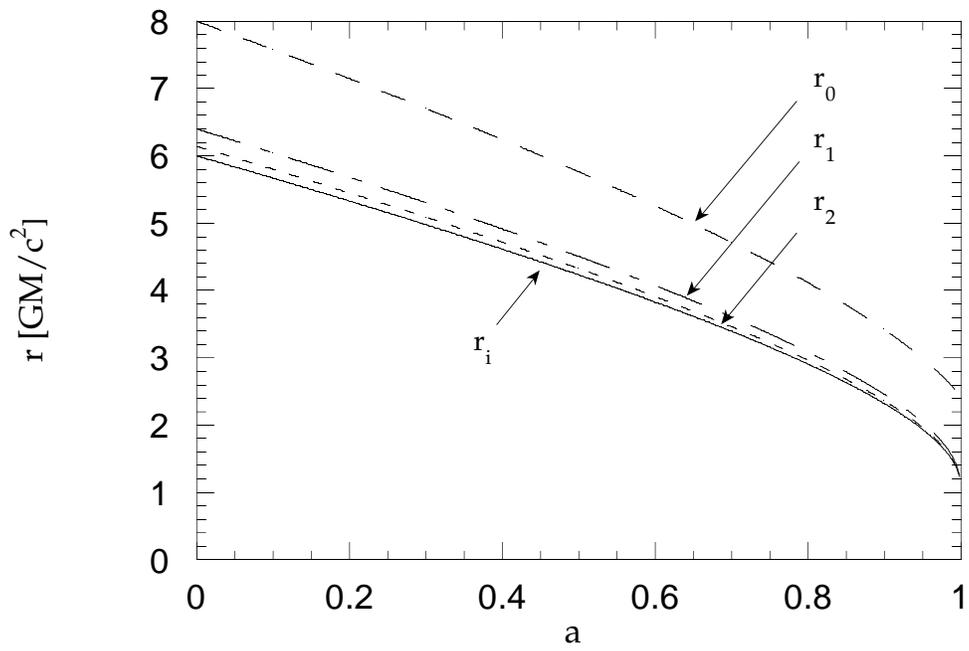
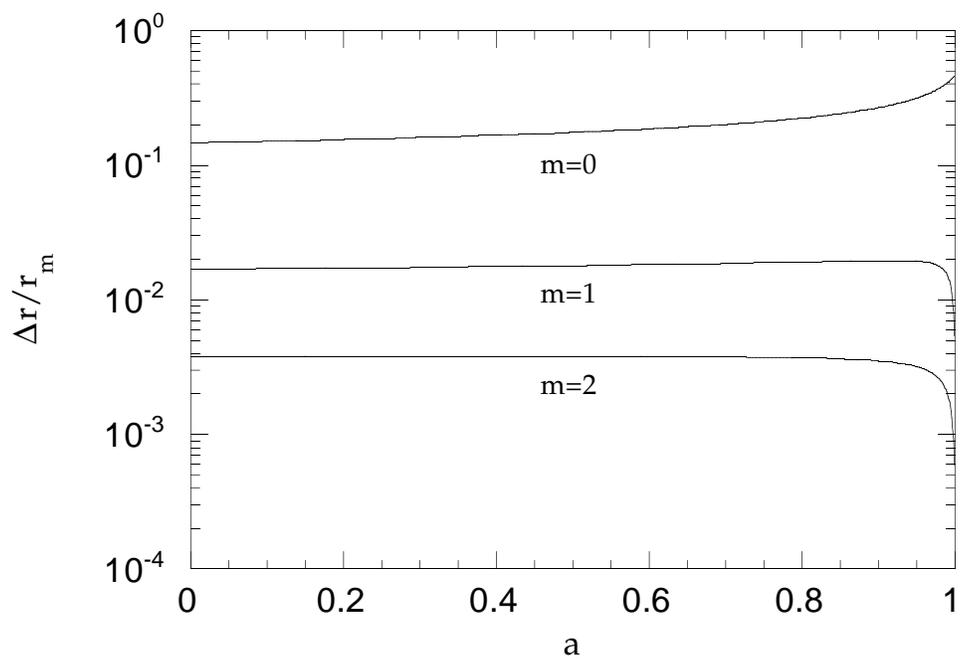
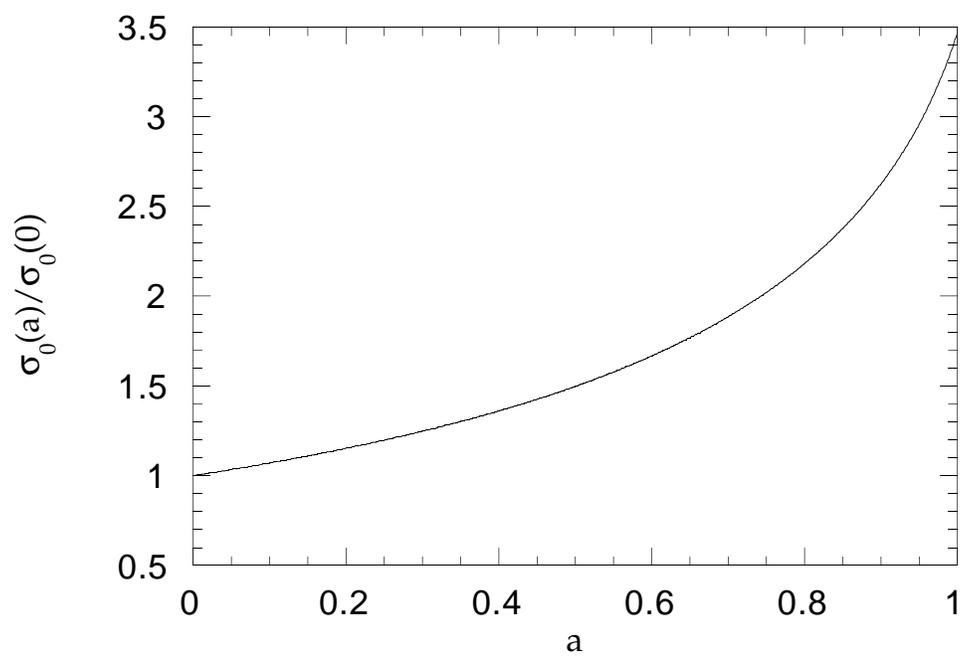
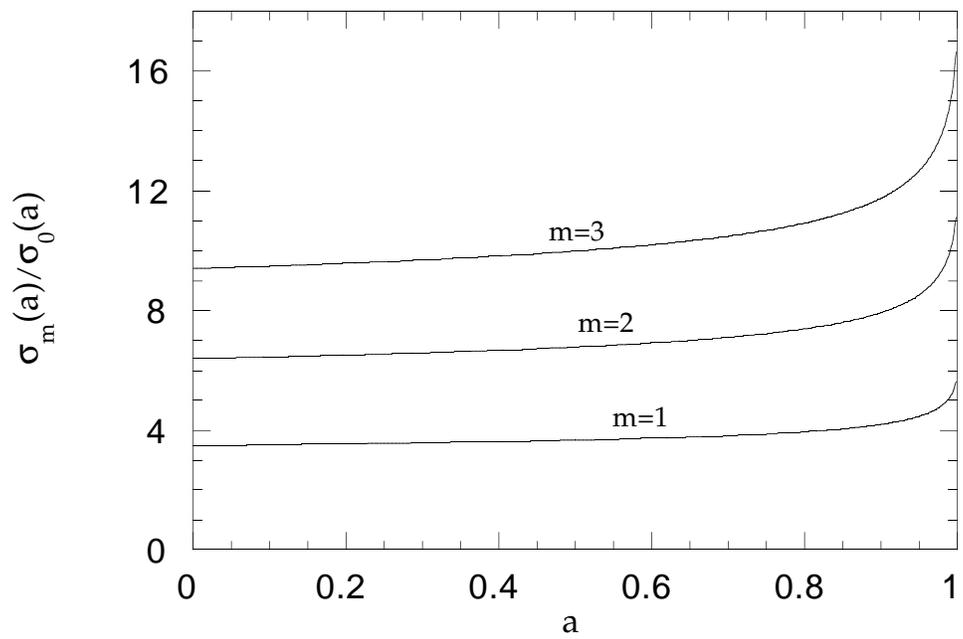
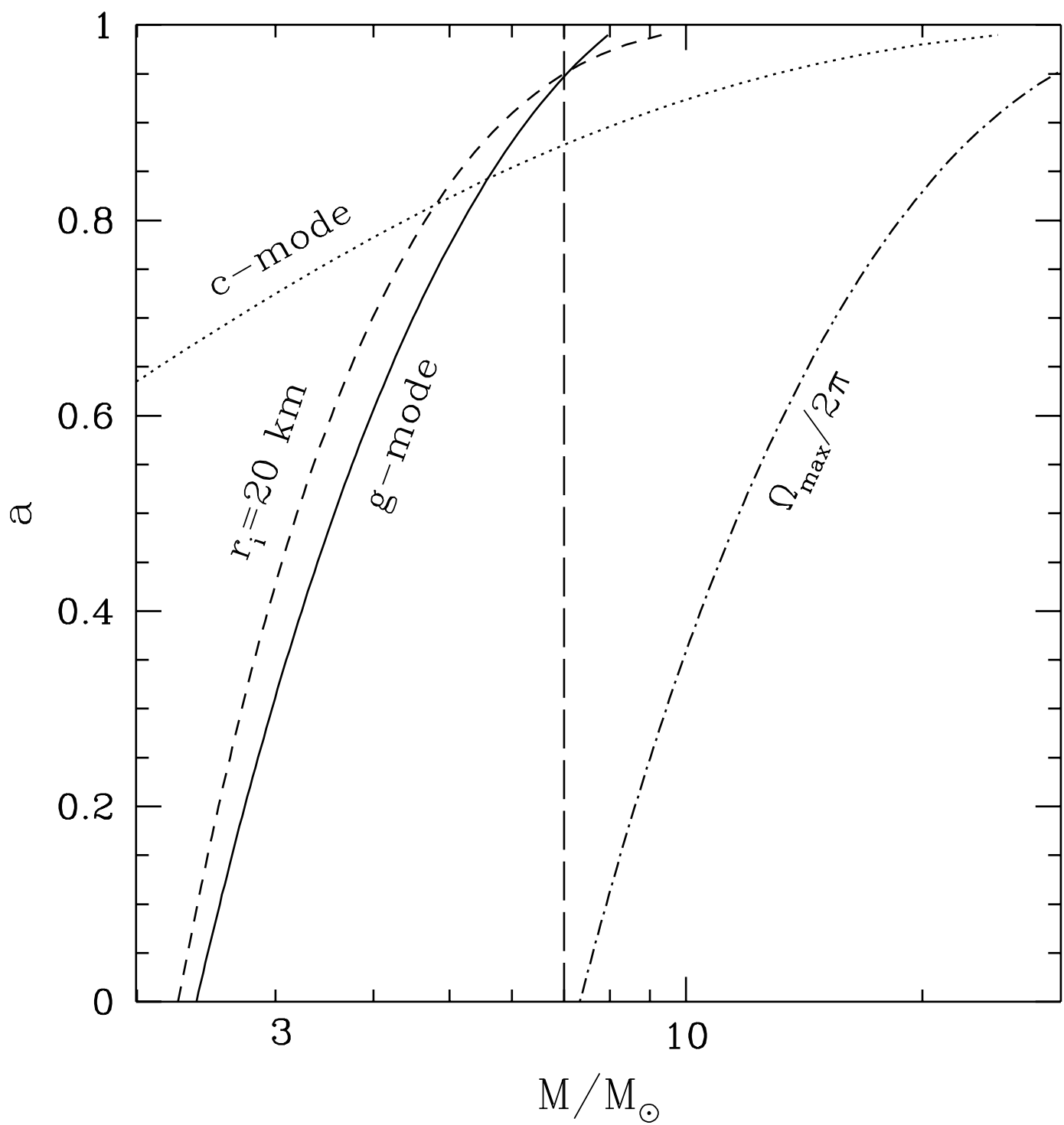
\end{document}